\documentclass[aps,prd,preprintnumbers,superscriptaddress]{revtex4-2}
\usepackage{graphicx} 
\usepackage{amsmath}
\usepackage{amssymb}
\usepackage{amsthm}
\usepackage{todonotes}
\usepackage{mathtools}
\usepackage{mathrsfs}
\usepackage{bm}
\usepackage{slashed} 
\usepackage{graphicx}
\usepackage{multirow}
\usepackage{tikz}
\usepackage[caption=false]{subfig}
\usepackage{relsize}	 
\usepackage{array}
\usepackage{float}
\usepackage{color}
\usepackage{xcolor}
\usepackage{soul}
\usepackage{verbatim} 
\usepackage{hyperref}
\allowdisplaybreaks
\newcommand{\beq}{\begin{equation}}
	\newcommand{\eeq}{\end{equation}}
\newcommand{\be}{\begin{eqnarray}}
	\newcommand{\ee}{\end{eqnarray}}

\newcommand{\Q}{{\scriptsize Q}}
\newcommand{\G}{{\scriptsize G}}
\newcommand{\bsq}{{\boldsymbol q}^{\perp}}
\newcommand{\bsqq}{{q}^{\perp 2}}

\newcommand{\bsk}{{\boldsymbol k}^{\perp}}
\newcommand{\bska}{{\boldsymbol\kappa}^\perp}

\newcommand{\bskasq}{{\kappa}^{\perp2}}

\newcommand{\bsb}{{\boldsymbol b}^{\perp}}

\newcommand{\bsP}{{\boldsymbol P}^{\perp}}

\newcommand{\es}{&=&}
\newcommand{\ps}{&+&}

\newcommand{\nn}{\nonumber}
\newcommand{\nnn}{\nonumber\\}

\begin{document}

\title{Sum rules for the Gravitational Form Factors using light-front dressed quark state}
	\author{Jai More}
	\email{Presented by Jai More at the International Conference on the Structure of Baryons, November 7–11th, Universidad Pablo de Olavide, Sevilla, Spain\\
	jai.more@iitb.ac.in} \affiliation{ Department of Physics,
		Indian Institute of Technology Bombay, Powai, Mumbai 400076,
		India}
	
	\author{Asmita Mukherjee}
	\email{asmita@phy.iitb.ac.in} \affiliation{ Department of Physics,
		Indian Institute of Technology Bombay, Powai, Mumbai 400076,
		India}
	
	\author{Sreeraj Nair}
	\email{sreeraj@impcas.ac.cn} \affiliation{Institute of Modern Physics, Chinese Academy of Sciences, Lanzhou 730000, China}
	\affiliation{School of Nuclear Science and Technology, University of Chinese Academy of Sciences, Beijing 100049, China}
	\affiliation{CAS Key Laboratory of High Precision Nuclear Spectroscopy, Institute of Modern Physics, Chinese Academy of Sciences, Lanzhou 730000, China}
	
	\author{Sudeep Saha}
	\email{sudeepsaha@iitb.ac.in} \affiliation{ Department of Physics,
		Indian Institute of Technology Bombay, Powai, Mumbai 400076,
		India}
\begin{abstract}
We consider a light-front dressed quark state, per se, instead of a proton state, we consider a simple composite spin-1/2 state of a quark dressed with a gluon. This perturbative model incorporates gluonic degrees of freedom, which enable us to evaluate the gravitational form factors (GFFs) of the quark as well as the gluon in this model \cite{More:2021stk, More:2023pcy}. We employ the Hamiltonian framework and choose the light-front gauge $A^+=0$. We calculate the four GFFs and corroborate the sum rules that GFFs satisfy. The GFF $D$ is attributed to information like pressure, shear, and energy distributions. We analyze some of these distributions for a dressed quark state at one loop in QCD.
\end{abstract}
	
\maketitle
\section{Introduction}
Despite the fact that the proton was discovered over a century ago, understanding its structure remains one of the most pressing questions in the field of hadron physics. Significant efforts have been made on both the theoretical and experimental fronts to shed light on issues such as the decomposition of mass and spin in terms of the underlying quarks and gluons that make up the proton. However, the strong force that holds these particles together presents a significant challenge to experimental study, making it difficult to obtain a complete picture of the proton's structure. Nonetheless, ongoing research efforts in high-energy collisions \cite{dHose2016,Kumeri_ki_2016, ATLAS:2018zdn, CLAS:2020yqf, JeffersonLabHallA:2022pnx,AbdulKhalek:2021gbh} and theoretical models studies \cite{Lorce:2018egm, Chakrabarti:2020kdc,Neubelt:2019sou,Hagler:2003jd,QCDSF-UKQCD:2007gdl,Rajan:2018zzy,Alexandrou:2020sml} hold promising for advancing our understanding. 

Understanding the energy-momentum tensor (EMT) of the hadronic matrix element would provide valuable information about the sum rules and gravitational coupling of quarks and gluons. The matrix element of the EMT encodes information about the gravitational form factors. 
These GFFs have become a focus of interest as they offer insights into the mechanical properties of nucleons, particularly through the additional form factor $D(Q^2)$, also known as the ``D-term". 

In this work, we study the quark and gluon GFFs in a theoretical framework that describes a relativistic spin-half system, specifically a quark dressed with a gluon at one loop. To accomplish this, we utilize the light-front Hamiltonian approach and expand the dressed quark state in Fock space using light-front wave functions (LFWFs). By analytically computing the two-particle quark-gluon LFWF using the light-front QCD Hamiltonian, we obtain the necessary overlap expressions using a two-component representation \cite{Zhang:1993dd}. Our approach builds upon previous research that utilized a similar model to investigate the Wigner function \cite{More:2017zqq,More:2017zqp}.

The manuscript is organized in the following manner:
In Sec.\ref{EMT} we start by writing the EMT of QCD and briefly discuss the dressed quark state.
 We give the final expressions for the quark and gluon GFFs. We then give the procedure to extract the GFFs for these states.
In Sec. \ref{Gff} we discuss the numerical plots of all the GFFs of the quark and the gluon. 
In Sec. \ref{Mechprop} we use the $D-$term to analyze the mechanical properties like pressure and shear distributions of the quark, the gluon and the total contribution. Finally, we conclude in Sec. \ref{conclusion}.
\section{Energy momentum tensor of QCD}\label{EMT}
\subsection{Definition}
The symmetric QCD EMT is defined as \cite{Harindranath:1998fm}, 
	\be
	\theta^{\mu \nu} \es \theta^{\mu \nu}_{\Q}+\theta^{\mu \nu}_\G,\\
\text{where}~~~~~\theta^{\mu \nu}_{\Q}\es 
	\frac{1}{2}\overline{\psi}\ i\left[\gamma^{\mu}D^{\nu}+\gamma^{\nu}D^{\mu}\right]\psi - g^{\mu \nu} \overline{\psi} \left(
	i\gamma^{\lambda}D_{\lambda} -m
	\right)
	\psi \label{emtqcd},\\
	\theta^{\mu \nu}_\G\es - F^{\mu \lambda a}F_{\lambda a}^{\nu} + \frac{1}{4} g^{\mu \nu} \left( F_{\lambda \sigma a}\right)^2 .
	\ee
	
	  The last term in Eq.~\ref{emtqcd} vanishes because of the equation of motion.
   	The EMT matrix element can be parameterized for a spin $-1/2$
system as follows \cite{Ji:2012vj}:
\be
	\langle P^{\prime}, S^{\prime}|	\theta^{\mu\nu}_i(0)|P,S \rangle \es\overline{U} (P^{\prime}, S^{\prime})\bigg[-B_i(q^2)\frac{\overline{P}^{\mu}\ \overline{P}^{ \nu}}{m}+\left(A_i(q^2)+B_i(q^2)\right)\frac{1}{2}(\gamma^{\mu}\overline{P}^{\nu}+\gamma^{\nu}\overline{P}^{\mu})\nnn
	\ps C_i(q^2)\frac{q^{\mu}q^{\nu}-q^2g^{\mu\nu}}{m}+\overline{C}_i(q^2)m\ g^{\mu\nu}\bigg]U(P, S), 
	\label{FF}
	\ee
	where  $\overline{P}^{\mu}=\frac{1}{2}(P^{\prime}+P)^{\mu}$ is the average nucleon four-momentum, $\overline{U}(P', S')$, $U(P, S)$ are the Dirac spinors for the state, and $m$ is the mass of the target state.  The Lorentz indices $(\mu, \nu)\ \equiv \{+,-,1,2\}$,  $i\equiv(Q, G)$. The quantities $A_i, B_i, C_i$  and  $\overline{C_i}$ are the quark or gluon gravitational form factors. 

\subsection{Dressed quark state}
 The light-front Hamiltonian approach provides a powerful framework for expanding the Fock state of any system with momentum $P$ and helicity $\lambda$ in terms of the light-front wave functions (LFWFs).  For instance, here we consider a dressed quark state that incorporates one gluon at one loop level, in which case we truncate the Fock space expansion at the two-particle state. 
 The state can be written as \cite{Harindranath:1998pd, Harindranath:1998fm}
	\be \label{state}
	|P,\lambda \rangle 
	\es \psi_1 (P, \lambda) b^{\dagger}_{\lambda}(P)|0 \rangle + \sum_{\lambda_1, \lambda_2}\int[k_1][k_2]   \sqrt{2(2\pi)^3P^+} \delta^3 (P-k_1-k_2)\ \psi_2 (P,\lambda|k_1,\lambda_1;k_2,\lambda_2) b^{\dagger}_{\lambda_1}(k_1) a^{\dagger}_{\lambda_2}(k_2)|0\rangle,  \nnn 
	\text{where}~~ [k]:\es \frac{dk^+ d^2\bsk} {\sqrt{16 \pi^3k^{+}}}.
   \label{state}
	\ee
 $b^\dagger$( $a^\dagger$) is the creation operator of quark (gluon). 
$\psi_1(P, \lambda)$ corresponds to a single particle wavefunction which contributes when $x=1$ and gives the normalization of the state. The two-particle LFWF, $\psi_2(P,\lambda|k_1,\lambda_1;k_2,\lambda_2)$ is related to the probability amplitude of finding a bare quark and a bare gluon with momentum (helicity) $k_1(\lambda_1)$ and $k_2(\lambda_2)$, respectively,  inside the target state.
 \subsection{Two-component formalism}
The boost invariant LFWF and two-particle LFWF are related as:
\be
\phi^{\lambda a}_{\lambda_1,\lambda_2}(x_i,\bska_i)\es\sqrt{P^+}\psi_2 (P,\lambda|k_1,\lambda_1;k_2,\lambda_2).
\ee
The constituent momenta in terms of the relative momenta ($x_i$, $\bska_i$) are   such that they satisfy the relation $x_1+x_2=1$ and $\bska_1+\bska_2=0$.
	\be
	k_i^+=x_iP^+,~~~~ \bsk_i=\bska_i+x_i \bsP,
	\ee 
	where $x_i$ is the longitudinal momentum fraction for the quark or gluon, inside the two-particle LFWF.
	The boost invariant two-particle LFWF can be written as
	\be\label{BILFWF}
	\phi^{\lambda a}_{\lambda_1,\lambda_2}(x,\bska)\es \frac{g}{\sqrt{2(2\pi)^3}}
\bigg[\frac{x(1-x)}{\bskasq+m^2x^2}\bigg]\frac{T^a}{\sqrt{x}} \chi_{\lambda_1}^{\dagger}\nnn
&\times&\bigg[\frac{2(\bska\cdot \boldsymbol{\epsilon}_{\lambda_2}^{\perp*})}{x}+\frac{1}{1-x}(\tilde{\sigma}^{\perp}\cdot\bska)(\tilde{\sigma}^{\perp}\cdot \boldsymbol{\epsilon}_{\lambda_2}^{\perp*})+im(\tilde{\sigma}^{\perp}\cdot \boldsymbol{\epsilon}_{\lambda_2}^{\perp*})\frac{x}{1-x}\bigg]\chi_{\lambda} \psi_1^{\lambda},
	\ee
We work in light-cone gauge $A^+=0$ and utilize the two-component formalism proposed in Ref \cite{Zhang:1993dd}. Here, $\chi$ represents the two-component spinor, $m$ denotes the mass of the quark, $T^a$ is the color SU(3) matrices and $\epsilon^\perp_{\lambda_2}$ is the polarization vector of the gluon. $\sigma^\perp (\perp\equiv 1,2)$ are Pauli matrices.

In this gauge, the quark field can be decomposed into:
	\begin{align}
		\psi_+= \begin{bmatrix}
			\xi\\0
		\end{bmatrix}, ~~~~\psi_-=\begin{bmatrix}
			0\\ \eta
		\end{bmatrix}, 
	\end{align}
	where the two-component quark fields are given by 
	\be
	\xi(y) \es \sum_{\lambda}\chi_{\lambda}\int \frac{[k]}{\sqrt{2(2\pi)^3}}[b_{\lambda}(k)e^{-ik\cdot y}+d^{\dagger}_{-\lambda}(k)e^{ik\cdot y}],
	\\
	\eta(y) \es \left(\frac{1}{i\partial^+}\right)\left[\sigma^{\perp}\cdot\left(i\partial^{\perp}+g A^{\perp}(y)\right)+im\right]\xi(y),
	\ee
 	$\eta(y)$ is the constrained field, which depends on $\xi(y)$, so it can be eliminated using the above equation. The dynamical components of the gluon field are given by 
	\begin{align}
		A^{\perp}(y)= \sum_{\lambda} \int \frac{[k]}{\sqrt{2(2\pi)^3k^+}}[{\bf\epsilon}^{\perp}_{\lambda}a_{\lambda}(k)e^{-i k \cdot y}+ {\bf \epsilon}^{\perp*}_{\lambda}a^{\dagger}_{\lambda}(k)e^{i k \cdot y}].
	\end{align}
	Here we have suppressed the colour indices.
 \subsection{The kinematical variables}
	The four momenta in light-front coordinates are defined as 
	\be
	P^\mu \es(P^+,\bsP,P^-),
	\ee
 where $P^-$ is the light-front energy, $P^+$ is the longitudinal momentum and $P^\perp$ is the transverse momentum. 
 We use the Drell-Yan frame (DYF), thus, momentum transfer is purely in the transverse direction, and $q^+=0$, Therefore, the four momenta of the initial and the final state are given by : 
	\be\label{initialmom}
	P^{\mu} \es\bigg(P^+, {\bf0}^{\perp}, \ \frac{m^2}{P^+}\bigg),\\
	\label{finalmom}
	P^{\prime\mu}\es\bigg(P^+,\ \bsq,\ \frac{\bsqq  + m^2}{P^+}\bigg), 
	\ee
	and the invariant momentum transfer
	\be\label{momtranfer}
	q^\mu\es(P^{\prime}-P)^\mu=\bigg(0, \ \bsq,   \frac{\bsqq}{P^+}\bigg).
	\ee
Hence, one can also write $q^2=-\bsqq$.
\subsection{Prescription to extract GFFs}
To calculate the GFFs, we sandwich the EMT between the dressed quark state given in Eq. (\ref{state}). This yields a generic form of the matrix element which can be expressed as:
	\be \label{matrixelement}
	\mathcal{O}^{\mu \nu }_{SS'} = \frac{1}{2}\left[\langle P'  ,S'|	\theta^{\mu \nu }_i(0)|P,S \rangle \right],
	\ee
where 
 $(S, S') \equiv \{ \uparrow,\downarrow \}$  is the helicity of the initial and final state. $\uparrow(\downarrow)$ positive (negative) spin projection along $z-$ axis.
 Then by selecting the appropriate Lorentz index ($\mu, \nu$) and by choosing the correct spin state, we can extract the GFFs for quark and gluon. A detailed description of the extraction of each GFFs can be seen in Ref \cite{More:2021stk, More:2023pcy}
\subsection{Quark and gluon GFFs}
The final expression for the four independent GFFs are as follows \cite{More:2021stk}:
\be\label{gffAexp}
A_{\Q}(q^2)\es 1+ \frac{g^2\ C_F}{2 \pi ^2}\left[\frac{11}{10} -\frac{4}{5}\left(1+\frac{2 m^2}{q^2}\right) \frac{f_2}{f_1}-\frac{1}{3} \log \left(\frac{\Lambda ^2}{m^2}\right)\right]\\
B_\Q(q^2)\es \frac{g^2 C_F}{12\pi^2}\ \frac{m^2}{q^2}\ \frac{f_2}{f_1},\label{gffBexp}\\
D_\Q(q^2)\es  \frac{5 g^2 C_F}{6\pi^2}\  \frac{m^2}{q^2} \ \bigg(1-f_1f_2\bigg)= 4 \ C_\Q(q^2),\label{gffDexp}\\
\overline{C}_\Q(q^2)\es \frac{g^2 C_F}{72\pi^2} \ \left(29-30 \ f_1 \ f_2+3~\mathrm{log}\bigg(\frac{\Lambda^2}{m^2}\bigg)\right),
\label{gffCbarexp}\ee
where
\be
f_1 :\es\frac{1}{2}\sqrt{1+\frac{4 m^2}{q^2}}, \\ f_2:\es\log\left(1+\frac{q^2\left(1+2f_1\right)}{2 m^2}\right) .
\ee
$C_F$ is the colour factor and $\Lambda$ is the ultra-violet cut-off.

The final expression for the gluon GFFs are as follows \cite{More:2023pcy}: 
\be
 A_\G(q^2)\!\!\!\!\es\!\!\! \frac{g^2C_F}{8\pi^2}\left[\frac{29}{9}+\frac{4}{3}\log\left(\frac{\Lambda^2}{m^2}\right)-\int dx\left(\left(1+\left(1-x\right)^2\right)+\frac{4m^2x^2}{q^2\left(1-x\right)}\right)\frac{\tilde{f_2}}{\tilde{f_1}}\right]\label{ag},\\
B_\G(q^2)\!\!\!\!\es\!\!\! -\frac{g^2C_F}{2\pi^2}\, \int dx\,\frac{m^2x^2}{q^2}\frac{ \tilde{f_2}}{ \tilde{f_1}}\label{bg},\\
D_\G(q^2) \!\!\!\!\es\!\!\! \frac{g^2C_F}{6\pi^2}\left[\frac{2 m^2}{3  q^2}+\int dx \frac{m^2}{ q^4 }\left(x\left(\left(2-x\right)q^2-4m^2x\right)\right)\right]\frac{\tilde{f_2}}{\tilde{f_1}}\label{dg},\\
\overline{C}_\G(q^2)\!\!\!\!\es\!\!\!   \frac{g^2C_F}{72\pi^2}\left[10 +9
\int dx~\left(x-\frac{4m^2x^2}{q^2\left(1-x\right)}\right)\frac{\tilde{f_2}}{\tilde{f_1}}-3\log\left(\frac{\Lambda^2}{m^2}\right)\right]\label{cbarg},
\ee
where, 
\be
&&\tilde{f_1}:= \sqrt{1+\frac{4m^2x^2}{q^2\left(1-x\right)^2}}.\\
&&\tilde{f_2}:= \log\left(\frac{1+\tilde{f_1}}{-1+\tilde{f_1}}\right).
\ee
$C_F$ is the colour factor and $\Lambda$ is the ultra-violet cut-off. The details of extraction and calculations of all the GFFs of quark and gluon are explained in the appendix of ref. \cite{More:2021stk,More:2023pcy}.
\section{Plots and numerical analysis of the quark and gluon GFFs}\label{Gff}
In this section, we study the analytical expressions of the quark and gluon GFFs, which were obtained in Eqs. \ref{gffAexp}-\ref{gffDexp} and \ref{ag} \ref{cbarg} respectively. To understand reasonably the behavior of GFFs, we plot the quark, the gluon and the total GFF as a function of the momentum transferred squared $(q^2)$ for each GFFs. For the calculation of both quark and gluon GFFs we use the following parameters: (1) the mass of the dressed quark  $m=0.3 \mathrm{GeV}$ (2) $g=C_F=1$ (3) the UV cutoff $\Lambda = 2~\mathrm{GeV}$.
\begin{figure}[ht]
\begin{minipage}{0.45\linewidth}
    \includegraphics[scale=0.4]{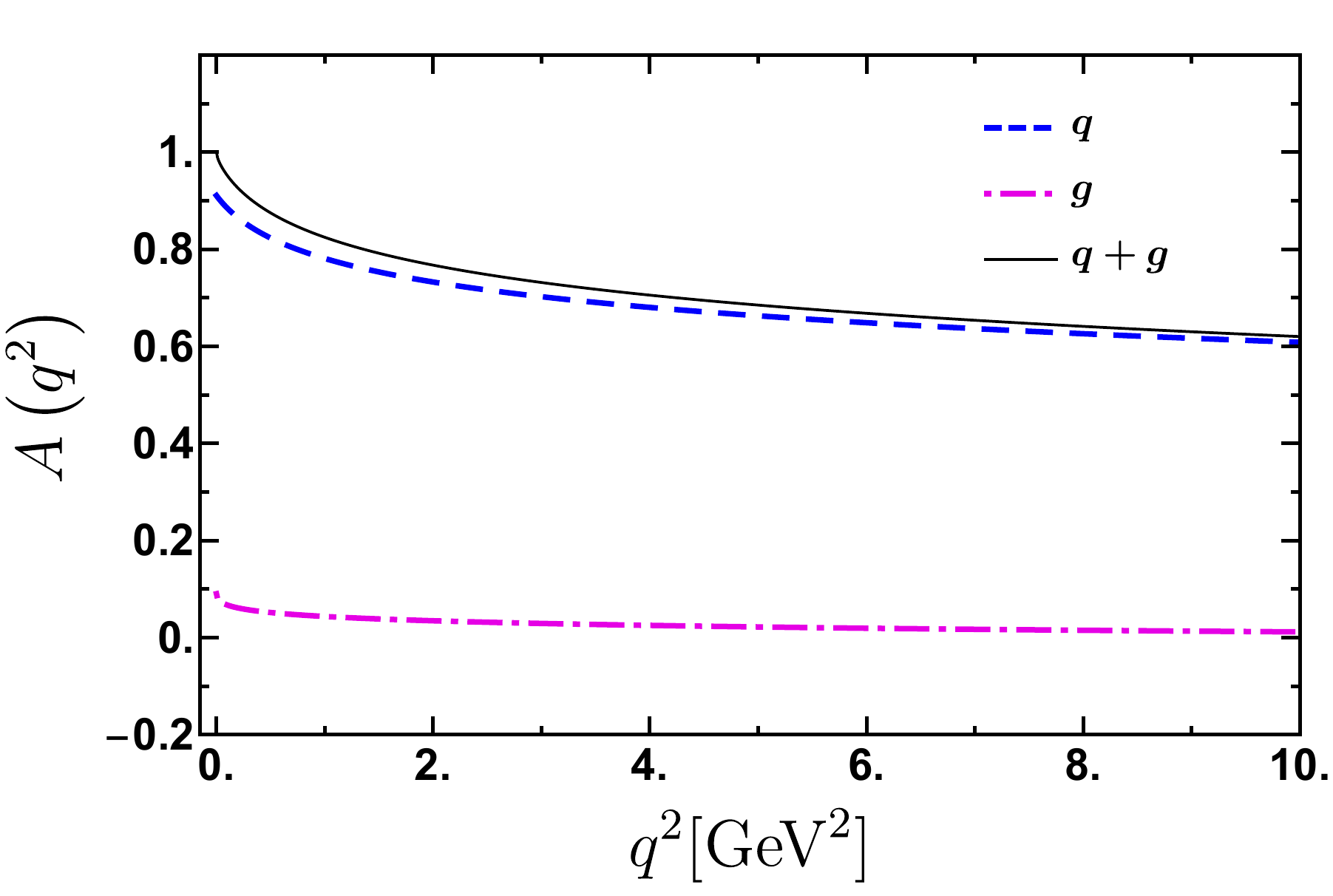}
\end{minipage}
\begin{minipage}{0.45\linewidth}
    \includegraphics[scale=0.4]{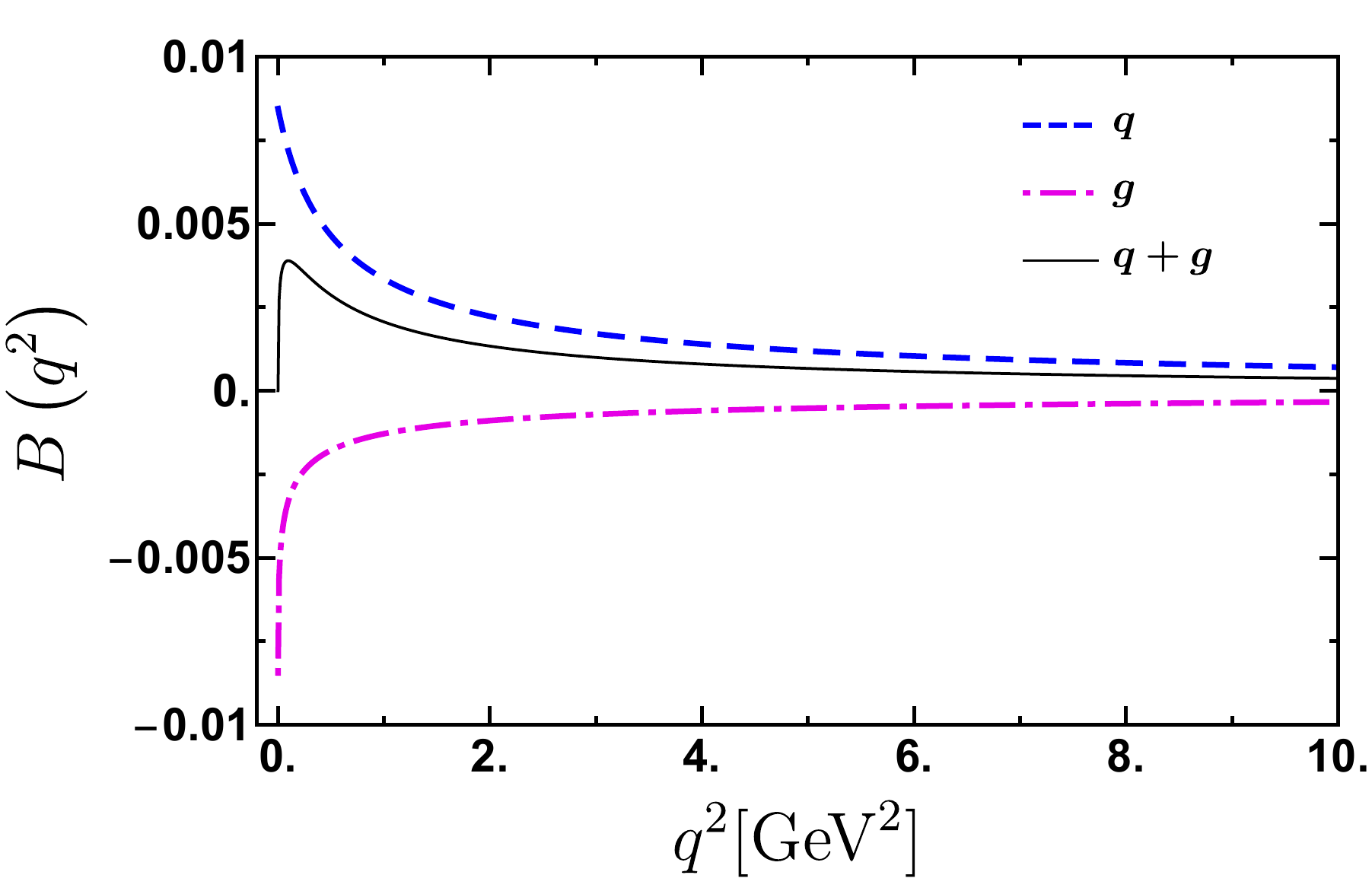}
\end{minipage}  
\caption{The GFF $A(q^2)$ and $B(q^2)$ is plotted as a function of $q^2$ for $m = 0.3 ~\mathrm{GeV}$ and $\Lambda = 2~\mathrm{GeV}$. The quark ($q$) and gluon ($g$) form factors are identified by the dashed blue curve and the dot-dashed magenta curve respectively. The solid black curve represents the sum of quark and gluon ($q+g$) contribution.}
\label{totalfiggffAnB}
\end{figure}
In Fig. \ref{totalfiggffAnB}, we plot the individual quark, the individual gluon and the total GFFs $A(q^2)$ and $B(q^2)$ as a functions of $q^2$. The GFF $A(q^2)$ for the quark and gluon depend on the cut-off $\Lambda$, but when we sum the quark and gluon contribution we obtain the total $A(q^2)$ which is independent of this cutoff. 
We observe, that the gluon contribution in the dressed quark state for the GFF $B(q^2)$ is negative. However, the quark contribution is positive. We observe that the total contribution to $B(q^2)$ at $q^2 \to 0$ vanishes.

{\it Some comparisons:} We set our parameter to QED limits and our results for $A(q^2)$ and $B(q^2)$ agree  with those reported in ref \cite{Brodsky:2000ii}. We also observe a qualitative agreement with the Lattice study of the gluon contribution to GFFs $A(q^2)$ and $B(q^2)$ of the nucleon, made in \cite{Deka:2013zha,Shanahan_2019}. 
\begin{figure}[ht]
\begin{minipage}{0.45\linewidth}
    \includegraphics[scale=0.4]{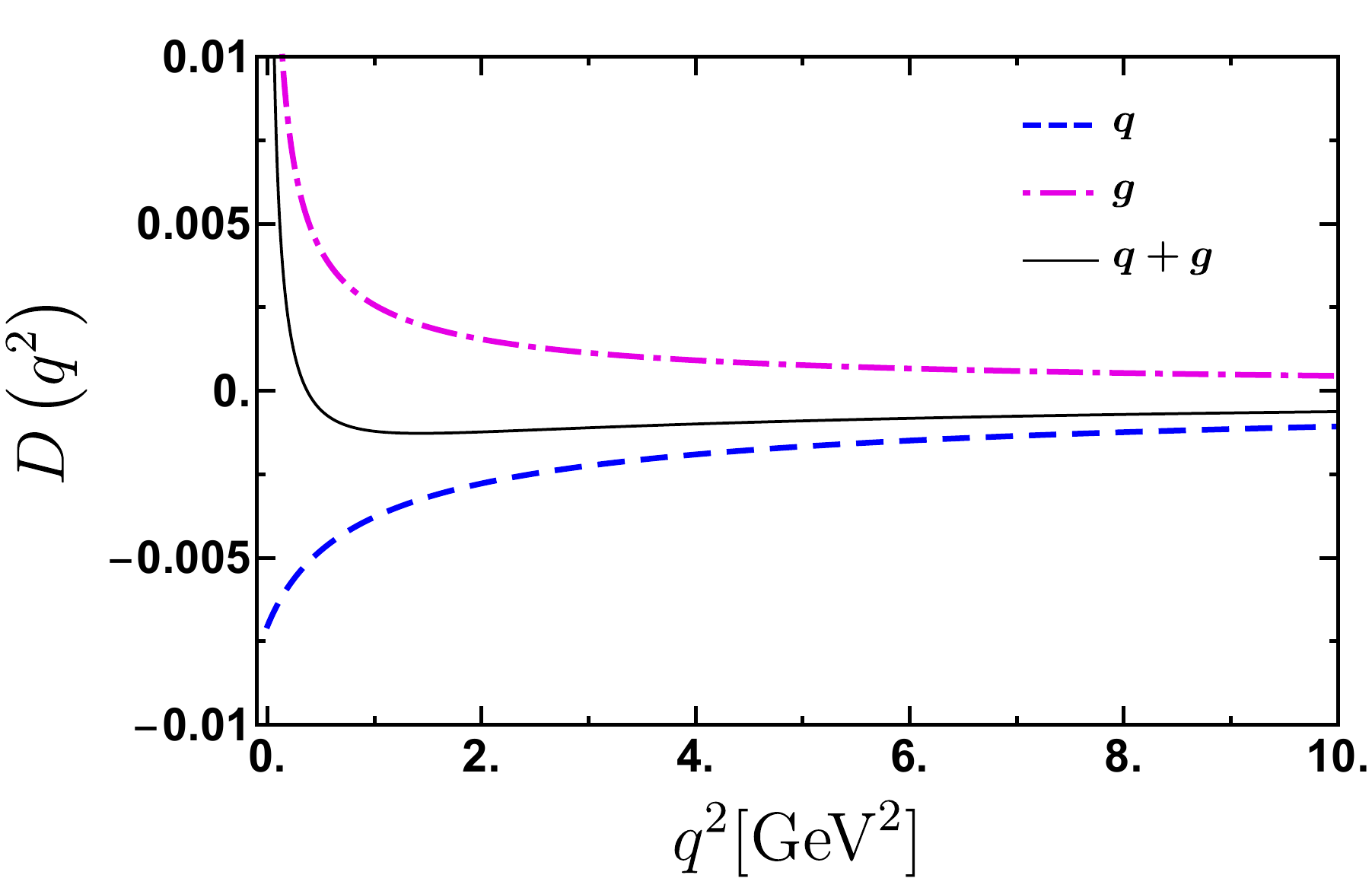}
\end{minipage}
\begin{minipage}{0.45\linewidth}
    \includegraphics[scale=0.4]{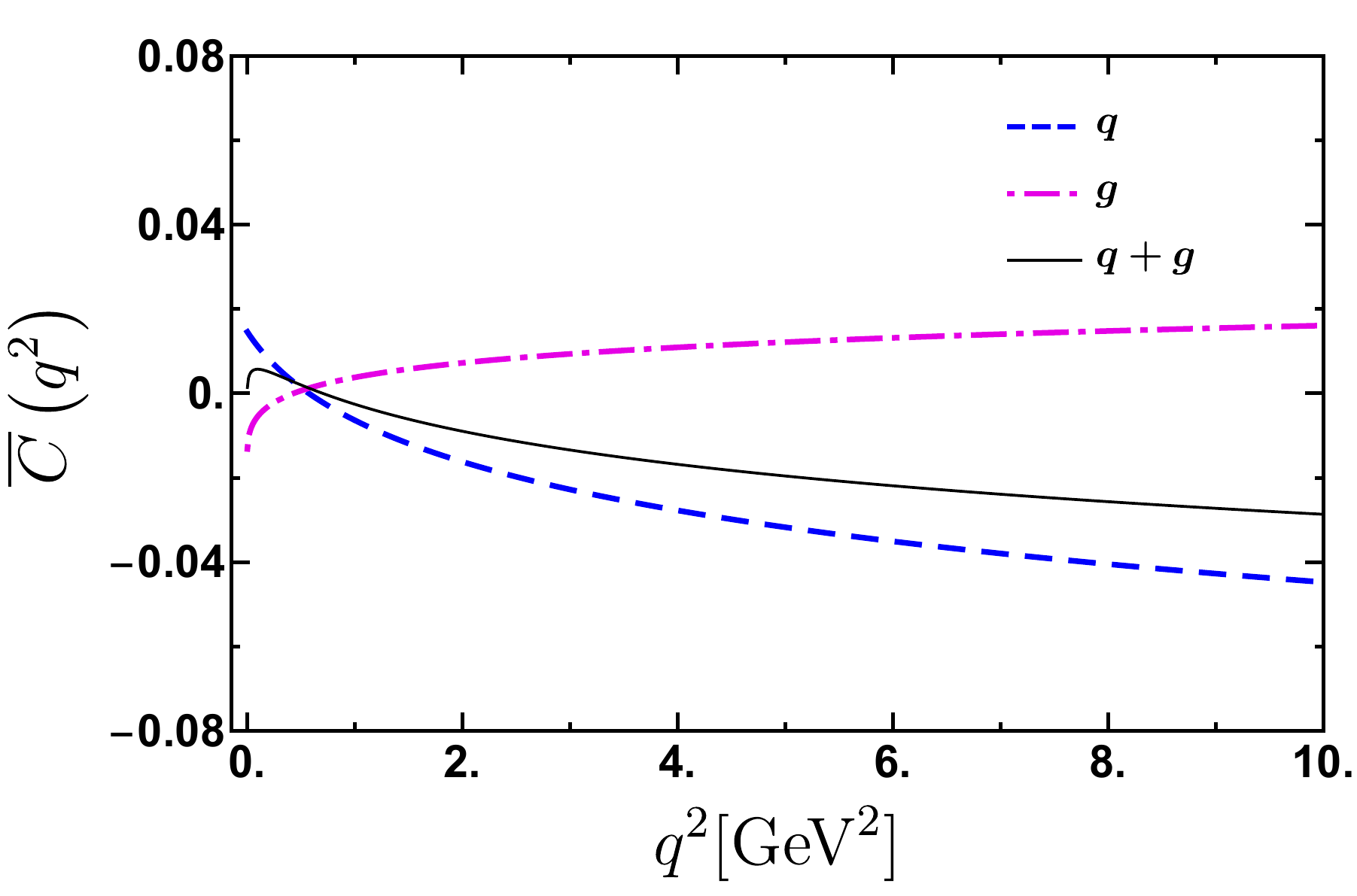}
\end{minipage}  
\caption{The GFF $D(q^2)$ and $\overline{C}(q^2)$ is plotted as a function of $q^2$ for $m = 0.3 ~\mathrm{GeV}$ and $\Lambda = 2~\mathrm{GeV}$. The dashed blue curve and the dot-dashed magenta curve show the quark ($q$) and gluon ($g$) form factors respectively. The quark and gluon ($q+g$) contribution is represented by the solid black curve.}
\label{totalfiggffDnCbar}
\end{figure}

In Fig. \ref{totalfiggffDnCbar}, we plot the individual quark, the individual gluon and the total GFFs $D(q^2)$ and $\overline{C}(q^2)$ as function of $q^2$. Here, we observe that $\overline{C}(q^2)$ for quark and gluon depends on the cut-off $\Lambda$, but the sum of quark and gluon contribution is independent of the cut-off as expected. Also, we observe that $\overline{C}(0)=0$ at $q^2=0$. We observe that $D(q^2)$ for quark is negative while for gluon it shows positive nature in the chosen range. As seen from the figure total $D(q^2)$ is negative for the entire range except for the region near zero.
\subsection{Sum rules of GFFs}
The constraint on GFF $A(q^2)$ corresponds to the momentum sum rule. For $B(q^2)$, the condition implies that the anomalous gravito-magnetic moment is zero for a spin$-1/2$ system \cite{Polyakov:2018zvc}. The {\it sum rule} we obtain for the form factors at $q^2 \to 0$ using the dressed quark state are:
\begin{itemize}
    \item $A(0)=1$, $B(0)=0$ \cite{ Lorce:2015lna,Lowdon_2017}
    \item $J(0)= \frac1{2} [A(0)+B(0)]$ \cite{PhysRevLett.78.610}
    \item $\overline{C}(0)=0$
\end{itemize}
 \section{Mechanical properties of dressed quark state}\label{Mechprop}
 In this section, we will focus on the pressure and shear distributions of a dressed quark state, specifically those related to the quark and gluon components. 
 A detailed discussion on topics like the force, the energy density and pressure distribution combinations for the quark and the gluon see ref\cite{More:2021stk, More:2023pcy}. 
 It is important to note that GFF $D$ provides a wealth of information about the pressure and shear distributions as highlighted in \cite{Polyakov:2018zvc}.
  The expressions for pressure and shear distributions in two dimensions \cite{Freese:2021czn} are 
\be
\label{Prefun}
p_i(b^{\perp}) \es \frac{1}{8m b^{\perp}} \frac{d}{db^{\perp}} \left[b^{\perp} \ \frac{d}{db^{\perp}} D_{i}(b^{\perp})\right]-m\ \overline{C}_i(b^{\perp}),\\
s_i(b^{\perp})\es-\frac{b^{\perp}}{4m} \frac{d}{db^{\perp }}\left[ \frac{1}{b^{\perp }} \frac{d}{db^{\perp }} D_i(b^{\perp})\right],
\label{Prefun2}
\ee

where
\be
\label{fgff}
F(b^{\perp})
\es  \frac{ 1}{(2\pi)^2}~\int d^2 \bsq \ e^{-i\bsq \bsb} \mathcal{F}(q^2) \nn \\
&=&\frac{1}{2\pi}\int_0^{\infty} d  q^{\perp} ~ q^{\perp} J_0\left( q^{\perp} b^{\perp}\right)\mathcal{F}(q^2),
\ee
		
where $\mathcal{F}=\left(A_i,B_i,D_i, \overline{C}_i\right)$, $i\equiv(Q, G)$. $J_0$ is Bessel's function of zeroth order. $m$ is the mass of the dressed quark state.
\begin{figure}[h]
\begin{minipage}{0.45\linewidth}
    \includegraphics[scale=0.4]{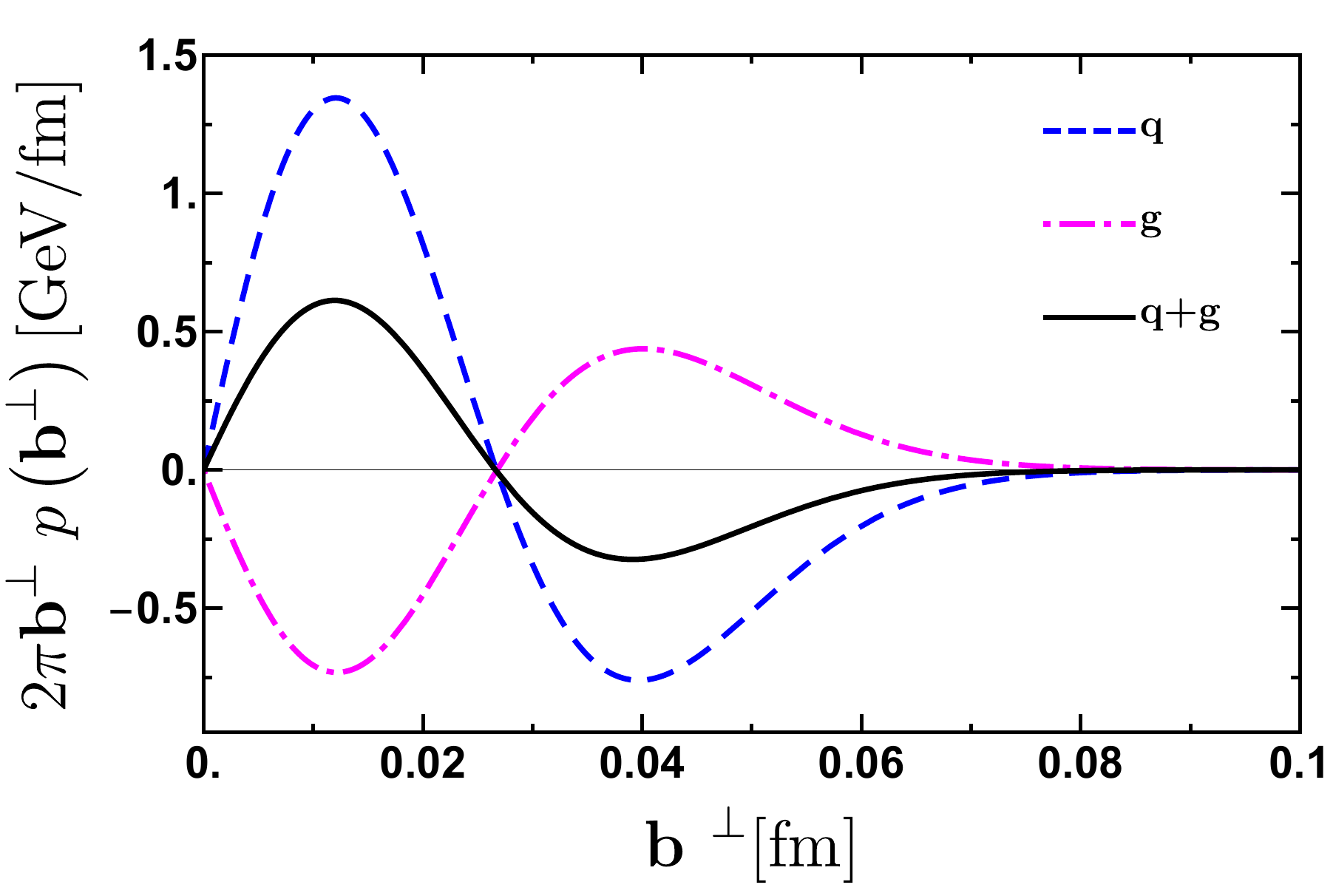}
\end{minipage}
\begin{minipage}{0.45\linewidth}
    \includegraphics[scale=0.4]{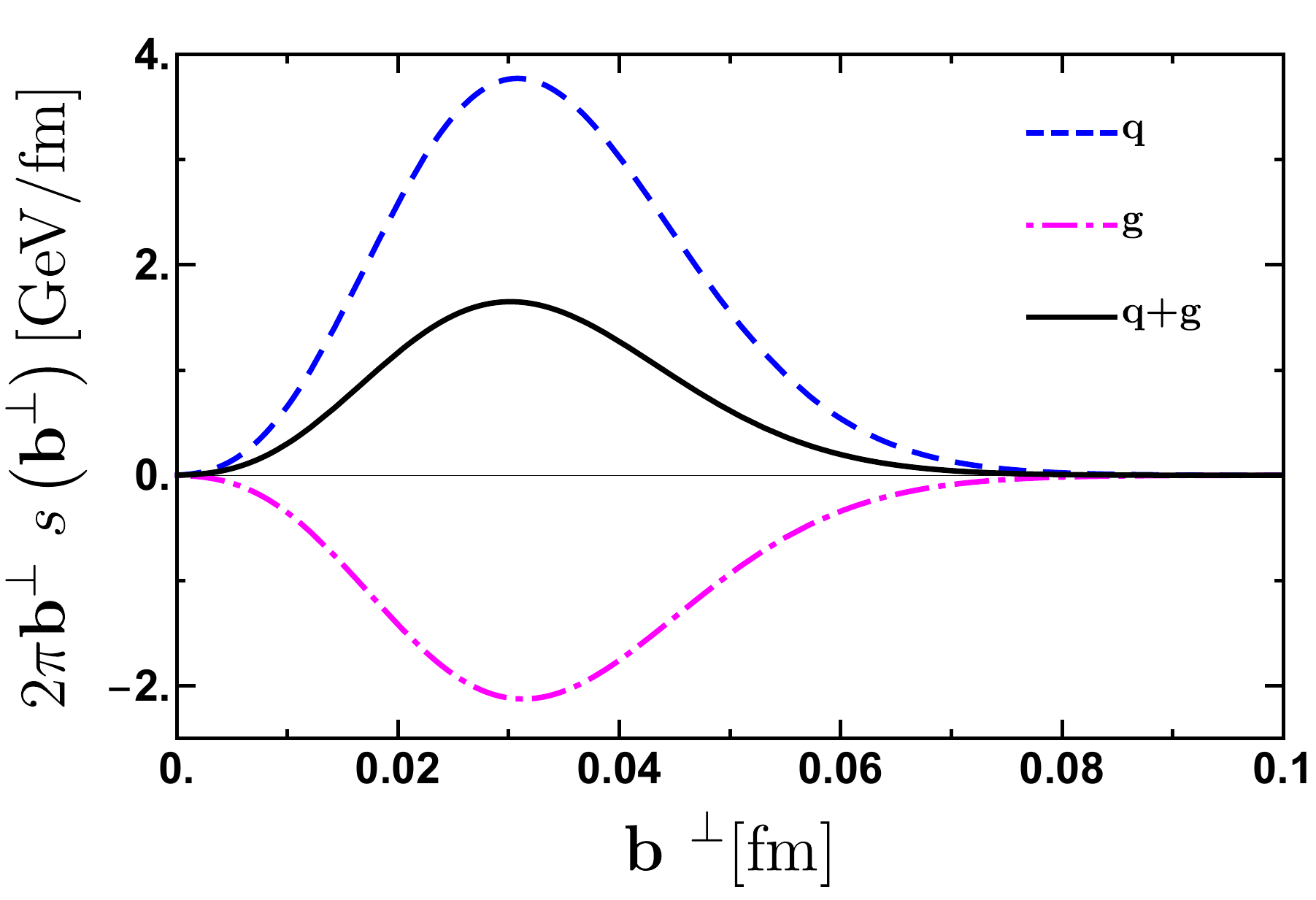}
\end{minipage}  
\caption{The pressure distribution and the shear force distribution are plotted as a function of $\bsb$ for 
 $\Delta=0.2$. The dashed blue curve and the dot-dashed magenta curve represent the quark and gluon contributions respectively. The sum of quark and gluon contribution is shown by the solid black curve.}\label{pns}
\end{figure}

In Fig. \ref{pns}, (left panel) we analyze the plot of $2 \pi \bsb ~p(\bsb)$ vs $q^2$ for the three distributions viz, the quark, the gluon and the sum of quark and gluon for $\Delta=0.2$. 
The quark and the gluon contribute complementary to each other and slightly differ in magnitude. The total pressure curve profile reveals that a positive core exists at the center of the two-particle system, while a negative pressure distribution occurs towards the outer region. This pressure distribution behavior is crucial for maintaining system stability, where the repulsive core is balanced by the confining pressure in the boundary region. Fig. \ref{pns}, (right panel) is a plot of $2 \pi \bsb ~s(\bsb)$ vs $q^2$. We observe that the quark contribution to shear force is positive, whereas the contribution of the gluon is negative. But, the total contribution to shear force is positive.  
\section{Summary}\label{conclusion}
The gravitational form factor of nucleons has been a topic of great interest in the theoretical side \cite{Polyakov:2019lbq,Neubelt:2019sou,Cebulla_2007,Chakrabarti:2015lba,Chakrabarti:2020kdc,Polyakov:2018exb}. In particular, the $D-$ term has been of prime importance due to its relation with pressure distribution inside the nucleons \cite{Lorce:2018egm,Metz:2021lqv, Freese:2021czn,Polyakov:2018zvc}. A lot of studies are dedicated to investigate the GFFs and mechanical properties of the bound-state systems. However, most of the phenomenological models do not incorporate gluons, thus information about the gluon cannot be perceived. In this work, instead of considering a proton state, we employ a light-front dressed quark state that takes into account gluonic degrees of freedom. Particularly, we examine a composite spin$-1/2$ state consisting of a quark dressed with a gluon. We have also evaluated mechanical properties like pressure distribution, shear distribution and 2-D energy distribution details of which can be found in refs. \cite{More:2021stk, More:2023pcy}.
This work provides a review of the quark and the gluon GFFs using the light-front dressed quark state. These GFFs have been shown to satisfy sum rules
that have been highlighted in this paper. Further, we investigate some of the mechanical properties like pressure and shear distributions of the quark and the gluon in this model. 
\section*{Acknowledgments}
J. M.  would like to thank the Department of Science and Technology (DST), Government of India, for financial support through grant No. SR/WOS-A/PM-6/2019(G) and Prof. Uma Sankar for partial travel support to attend the conference under the grant `PGRDFI94091'.
\section*{Declaration of competing interest}
The authors declare that they have no known competing financial interests or personal relationships that could have appeared to influence the work reported in this paper.
\bibliographystyle{elsarticle-num}
\bibliography{references}

\end{document}